# On the problem of projectile motion in a medium with quadratic resistance

## Peter Chudinov

*Department of Engineering, Perm State Agro-Technological University, 614990, Perm, Russia*

*E-mail: chupet@mail.ru*

A classic problem of the motion of a projectile thrown at an angle to the horizon in a medium with a quadratic resistance law is studied. An approximate analytical solution of the equations of projectile motion is presented, which has a high accuracy. The proposed formulas are universal, that is, they can be used for any initial conditions of throwing over a wide range of the change of medium resistance coefficient. The motion of a shuttlecock of badminton, bullet, table tennis ball and volleyball are presented as examples.

Keywords: Projectile motion; quadratic resistance law; parametric definition of the trajectory.

## 1. Introduction

The classical problem of the motion of a projectile thrown at an angle to the horizon in a medium with resistance still attracts the attention of researchers. An exact solution to the problem in motion with a quadratic law of medium resistance has not yet been found. Therefore, many different approximate analytical solutions are proposed using various approaches, including expansions in series [1 - 7]. Some approximate solutions use special functions, for example, the Lambert W function [2 – 3].

The solution of the problem in the form of power series has one significant problem. It lies in the fact that, generally speaking, it is necessary to verify the convergence of the obtained series. Here is what the authors of [5] note: " Providing a function in terms of a power series, or a ratio of power series, raises the question of radius of convergence. We have made several attempts to solve this issue by applying various techniques. Unfortunately, it was so far not possible to analytically determine such an expression for the convergence radius from the recursion formula, so that this is still an open problem". In paper [4] an analytic solution of the problem of two-dimensional projectile motion with quadratic resistance law for a large angle of projection is obtained using the homotopy analysis method. The Cartesian coordinates of the projectile $x, y$ are defined as the following time series:

$$x(t) = \sum_{m=0}^{N} \sum_{n=0}^{m} \frac{1}{n+1} a_{m,n} t^{n+1}, \quad y(t) = \sum_{m=0}^{N} \sum_{n=0}^{m} \frac{1}{n+1} b_{m,n} t^{n+1}. \qquad (1)$$

Further, the authors of [4] note that "this solution may be invalid for a large dimensionless initial velocity, because the present solution is a polynomial expression". The same circumstance is pointed out by the authors of the article [6]: "At the given differential equations and the boundary conditions, a series solution of the problem can always be obtained. A problem arises if the series obtained is a divergent one. For most of the reasonable initial conditions, the series obtained for the problem is divergent". Therefore, the approach to solving the problem in the form of power series is somewhat complicated and limited.

At the same time, the problem of the motion of a projectile thrown at an angle to the horizon is widely used for educational purposes. Introductory physics courses very often contain a solution to the problem in the simplest cases (in the absence of air resistance or with a linear law of medium resistance). The next logical step is to describe the motion of a projectile in a medium with a quadratic drag law using elementary functions. Therefore, solving the problem with the help of special functions is not very convenient in methodological and educational terms. It is more preferable to represent the solution of the problem using elementary functions that are well known to students.

These shortcomings can be avoided by an approach based on the direct integration of differential equations of motion. This idea was implemented in [1, 8]. In [8], an approximate analytical solution of this classical problem was obtained, which has high accuracy and is efficient under any initial conditions and for any values of the projectile parameters. In this study, the results of [8] are further developed. The purpose of the study is to test the applicability of the proposed formulas and compare them with existing solutions in the indicated ranges of initial conditions and parameters. The conditions of applicability of the quadratic resistance law are deemed to be fulfilled, i.e. Reynolds number $Re$ lies within $1 \times 10^3 < Re < 2 \times 10^5$ [4]. The Reynolds number is a dimensionless measure of the magnitude of inertia relative to that of viscous forces in a fluid flow. For a projectile of diameter $D$ thrown at speed $V_0$ in air, the Reynolds number is typically chosen (as in this paper) to be $V_0 D / \nu$, where $\nu$ is the kinematic viscosity of air. Magnus forces are not included in this work.

## 2. Equations of projectile motion

Here we state the formulation of the problem and the equations of the motion [8]. Let us consider the motion of a projectile with mass $m$ launched at an angle $\theta_0$ with an initial speed $V_0$ under the influence of the force of gravity and resistance force $R = mgkV^2$. Here $g$ is the acceleration of gravity, $k$ is the drag constant and $V$ is the speed of the object. Air resistance force $R$ is proportional to the square of the speed of the projectile and is directed opposite the velocity vector [see Fig. 1]. In ballistics, the movement of a projectile is often studied in projections on natural axes. The equations of the projectile motion in this case have the form

$$\frac{dV}{dt} = -g\sin\theta - gkV^2, \quad \frac{d\theta}{dt} = -\frac{g\cos\theta}{V}, \quad \frac{dx}{dt} = V\cos\theta, \quad \frac{dy}{dt} = V\sin\theta. \quad (2)$$

Here $\theta$ is the angle between the tangent to the trajectory of the projectile and the horizontal, $x, y$ are the Cartesian coordinates of the projectile. The well-known solution [9] of system (2) consists of an explicit analytical dependence of the velocity on the slope angle of the trajectory and three quadratures

$$V(\theta) = \frac{V_0 \cos\theta_0}{\cos\theta\sqrt{1 + kV_0^2 \cos^2\theta_0 \left(f(\theta_0) - f(\theta)\right)}}, \quad f(\theta) = \frac{\sin\theta}{\cos^2\theta} + \ln\tan\left(\frac{\theta}{2} + \frac{\pi}{4}\right),$$

$$x = x_0 - \frac{1}{g}\int_{\theta_0}^{\theta} V^2 d\theta, \quad y = y_0 - \frac{1}{g}\int_{\theta_0}^{\theta} V^2 \tan\theta d\theta, \quad t = t_0 - \frac{1}{g}\int_{\theta_0}^{\theta} \frac{V}{\cos\theta} d\theta. \quad (3)$$

Here $t_0$ is the initial value of the time, $x_0, y_0$ are the initial values of the coordinates of the projectile. In the following formulas, we take $x_0 = 0$, $t_0 = 0$, $y_0 \geq 0$. The drag coefficient $k$, used in formulas (2) – (3), can be calculated through the terminal speed: $k = 1/V_{term}^2$. Terminal velocity is the maximum velocity attainable by an object as it falls through a fluid (air is the most common example).

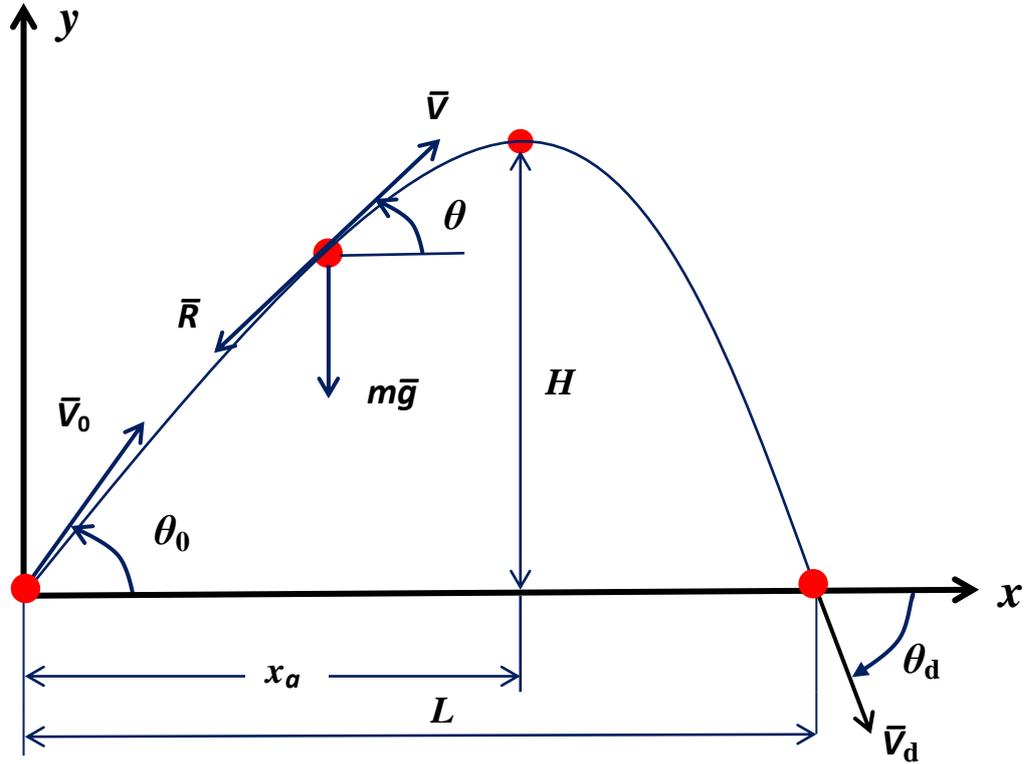

**Fig. 1.** Basic motion parameters.

## 3. Computational formulas of the problem

In most articles devoted to the problem under study, approximate analytical solutions are constructed in a form similar to relations (1), i.e. in the form of dependences of the Cartesian coordinates of the projectile on time: $x(t), y(t)$. In [8], an approximate analytical solution of the problem under study was also obtained. It was found by direct integration of quadratures (3) with the help of a special technique - approximation of subintegral expressions. This solution has a different form - the dependences of coordinates and time on the trajectory angle $\theta$ - $x(\theta), y(\theta), t(\theta)$. The solution of the projectile motion problem in the form of functions $x(\theta), y(\theta), t(\theta)$ has some advantage over the solution of the form $x(t), y(t), \theta(t)$. It lies in the fact that the value of the angle of inclination of the trajectory of the projectile is known in advance for two remarkable positions - the vertex of the trajectory (here $\theta = 0$) and movement along the asymptote (here $\theta = -\pi/2$). Substituting the specified values in the functions $x(\theta), y(\theta), t(\theta)$, we immediately find the following important characteristics of the movement: $x_a = x(0)-$ abscissa of the vertex of the projectile trajectory, $H = y(0)-$ maximum height of the trajectory, $t_a = t(0)-$ projectile lift time, $x_{as} = x(-\pi/2)-$ the value of the vertical asymptote of the projectile trajectory (see Fig. 1).

According to the approach in [8], we divide the entire range of the trajectory angle $\theta_0 \geq \theta > -\pi/2$ into three intervals: $\theta_0 \geq \theta \geq 0$, $0 \geq \theta \geq \theta_1$, $\theta_1 \geq \theta > -\pi/2$. The value $\theta_1$ is determined by the equality $\theta_1 = -\frac{\theta_0}{2} - \frac{\pi}{4}$. Such a partition allows one to construct a solution over the entire interval of angle change $\theta$. The first interval corresponds to the projectile lifting stage, the other two intervals correspond to the descent stage.

On the first interval, the functions $x(\theta), y(\theta), t(\theta)$ have the following form:

$$x_1(\theta) = x_0 + A_1 \arctan\left(\frac{b_3(\tan\theta - \tan\theta_0)}{\tan\theta + \tan\theta_0 + 2(b_2 \tan\theta \tan\theta_0 - b_1)}\right),$$

$$y_1(\theta) = y_0 + \frac{1}{2gk\alpha_2} \ln\left(\frac{b_2 \tan^2\theta + \tan\theta - b_1}{b_2 \tan^2\theta_0 + \tan\theta_0 - b_1}\right) - \frac{x_1(\theta)}{2b_2},$$

$$t_1(\theta) = t_0 + \frac{1}{g\sqrt{k\alpha_2}}\left[\arcsin\left(\frac{1 + 2b_2 \tan\theta_0}{\Delta_1}\right) - \arcsin\left(\frac{1 + 2b_2 \tan\theta}{\Delta_1}\right)\right].$$

On the second interval we have

$$x_2(\theta) = x_1(0) + A_2 \arctan\left(\frac{b_4}{1 - 2b_1 \cot\theta}\right),$$

$$y_2(\theta) = y_1(0) + \frac{x_2(\theta) - x_1(0)}{2b_2} + \frac{1}{2gk\alpha_2} \ln\left(\frac{b_1}{b_1 - \tan\theta + b_2 \tan^2\theta}\right),$$

$$t_2(\theta) = t_1(0) + \frac{1}{g\sqrt{-k\alpha_2}}\left[\arcsin\left(\frac{2b_2 \tan\theta - 1}{\Delta_2}\right) + \arcsin\left(\frac{1}{\Delta_2}\right)\right].$$

On the third interval we have

$$x_3(\theta) = x_2(\theta_1) + A_3\left[\arctan\left(\frac{1 - 2d_1 \tan\theta}{d_2}\right) + \arctan\left(\frac{2d_1 \tan\theta_1 - 1}{d_2}\right)\right],$$

$$y_3(\theta) = y_2(\theta_1) + \frac{x_3(\theta) - x_2(\theta_1)}{2d_1} + \frac{1}{2gk\beta_2} \ln\left(\frac{d_0 - \tan\theta_1 + d_1 \tan^2\theta_1}{d_0 - \tan\theta + d_1 \tan^2\theta}\right),$$

$$t_3(\theta) = t_2(\theta_1) + \frac{1}{g\sqrt{-k\beta_2}}\left[\arcsin\left(\frac{2d_1 \tan\theta - 1}{\Delta_3}\right) + \arcsin\left(\frac{1 - 2d_1 \tan\theta_1}{\Delta_3}\right)\right]. \tag{4}$$

Index value $i$ in functions $x_i(\theta), y_i(\theta), t_i(\theta)$ corresponds to the number of the movement interval $(i = 1, 2, 3)$. In relations (4), the following notation is introduced:

$$\alpha_1 = 2\cot\theta_0 \ln\tan\left(\frac{\theta_0}{2} + \frac{\pi}{4}\right), \quad \alpha_2 = \frac{1}{\sin\theta_0} - \frac{\alpha_1}{2}\cot\theta_0, \quad b_1 = \left(\frac{1}{kV_0^2 \cos^2\theta_0} + f(\theta_0)\right)/\alpha_1,$$

$$b_2 = \frac{\alpha_2}{\alpha_1}, \quad b_3 = \sqrt{-1 - 4b_1 b_2}, \quad b_4 = \sqrt{-1 + 4b_1 b_2}, \quad A_1 = \frac{2}{gk\alpha_1 b_3}, \quad A_2 = \frac{2}{gk\alpha_1 b_4},$$

$$\Delta_1 = \sqrt{1+4b_1b_2}, \quad \Delta_2 = \sqrt{1-4b_1b_2}, \quad \beta_2 = \frac{\left(f(\theta_1)+f(89°)\right)\cos\theta_1 - 2\left(\tan\theta_1 + \tan 89°\right)}{(\tan\theta_1 + \tan 89°)^2 \cos\theta_1},$$

$$\beta_1 = \frac{2(1+\beta_2\sin\theta_1)}{\cos\theta_1}, \quad \beta_0 = f(\theta_1) - \beta_1\tan\theta_1 + \beta_2\tan^2\theta_1, \quad d_0 = \frac{1}{\beta_1}\left(\frac{1}{kV_0^2\cos^2\theta_0} + f(\theta_0) - \beta_0\right),$$

$$d_1 = \frac{\beta_2}{\beta_1}, \quad d_2 = \sqrt{4d_0d_1 - 1}, \quad A_3 = \frac{2}{gk\beta_1 d_2}, \quad \Delta_3 = \sqrt{1-4d_0d_1}, \quad f(\theta) = \frac{\sin\theta}{\cos^2\theta} + \ln\tan\left(\frac{\theta}{2}+\frac{\pi}{4}\right).$$

Thus, on each of the three intervals, the movement of the projectile is described by the equations:

at $i=1$  $x = x_1(\theta),\; y = y_1(\theta),\; t = t_1(\theta)$;
at $i=2$  $x = x_2(\theta),\; y = y_2(\theta),\; t = t_2(\theta)$;
at $i=3$  $x = x_3(\theta),\; y = y_3(\theta),\; t = t_3(\theta)$.

Collectively, these equations describe the motion of the projectile over the entire interval of change in the trajectory angle $\theta_0 \geq \theta \geq -\pi/2$. If during the motion of the projectile the trajectory angle $\theta$ is within the limits $\theta_0 \geq \theta \geq \theta_1$, then the functions $x_3(\theta), y_3(\theta), t_3(\theta)$ are not used to describe of the movement. The projectile trajectory is given parametrically by the functions $x(\theta), y(\theta)$. Important motion characteristics $x_a, H, t_a$ are determined by direct substitution of the value $\theta = 0$ in functions (4).

Formulas (4) are an improved version of the previously obtained formulas [8]. At very small values of the drag coefficient ($k = 10^{-12}$), formulas (4) are transformed into formulas of the theory of parabolic projectile motion. The value $k = 0$ cannot be used in formulas (4), since division by zero occurs.

The value $\theta \cong -\pi/2$ corresponds to the asymptote movement. It is easy to obtain the numerical value of the asymptote $x_{as}$ by substituting the value $\theta = -\pi/2$ into the function $x_3(\theta)$:

$$x_{as} = x_3\left(-\frac{\pi}{2}\right) = A_1 \arctan\left(\frac{b_3}{2b_1\cot\theta_0 - 1}\right) + A_2 \arctan\left(\frac{b_4}{1 - 2b_1\cot\theta_1}\right) + A_3 \operatorname{arccot}\left(\frac{1 - 2d_1\tan\theta_1}{d_2}\right).$$
(5)

Multipliers $A_1, A_2, A_3$ and coefficients $b_1, b_3, b_4, d_1, d_2$ were introduced earlier. Note that the asymptote value is calculated directly from the initial conditions of motion $V_0, \theta_0$, without integrating the equations of motion of the projectile.

## 4. Calculation results

In order to check the applicability of formulas (4), the widest ranges of changes in the initial conditions of throwing $V_0, \theta_0$ and the drag coefficient $k$, which can only be found in the literature, were used in the calculations:

$$0 \leq k \leq 0.022 \text{ s}^2/\text{m}^2, \quad 49' \leq \theta_0 \leq 89°, \quad 0 < V_0 \leq 823 \text{ m/s}.$$

All the above figures show the trajectory of the projectile. The thick solid black lines are obtained by numerical integration of system (2) with the aid of the 4-th order Runge-Kutta method (RK4). The red dot lines are obtained using analytical formulas (4).

Fig. 2 shows the trajectory of the projectile with the following data [6]:

$$V_0 = 823 \text{ m/s}, \quad \theta_0 = 49', \quad k = 0.000107 \text{ s}^2/\text{m}^2, \quad g = 9.8 \text{ m/s}^2. \tag{6}$$

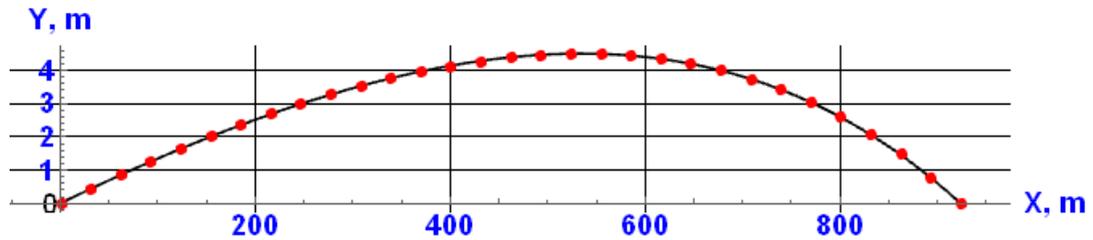

**Fig .2.** The trajectory of the projectile at a small angle of throw.

Fig. 2 shows that formulas (4) perfectly approximate the trajectory. The main characteristics of movement are given in Table 1. Here $L$ is the flight range, $T$ is the time of motion, $\theta_d$ is impact angle with respect to the horizontal (see Fig. 1).

The table demonstrates a remarkable agreement between the results of the numerical solution and the results of applying formulas (4).

**Table 1.** The main parameters of the motion.

|  | $L$, m | $T$, s | $H$, m | $x_a$, m | $t_a$, s | $\theta_d$, ° |
|---|---|---|---|---|---|---|
| RK4 | 923.65 | 1.894 | 4.50 | 534.17 | 0.870 | -1.53° |
| Formulas (4) | 923.65 | 1.894 | 4.50 | 534.17 | 0.870 | -1.53° |
| Das, Roy [6] | 923.40 | 1.894 | 4.59 | 534.00 | 0.870 | --- |

It should be noted here that in [6] it was used solution using a 12th-order time series and the Padé approximant, where the series coefficients were found numerically.

Fig. 3 shows the trajectory of a projectile with the same parameters (6) and with a non-zero height of the point of throw, considered in [3].

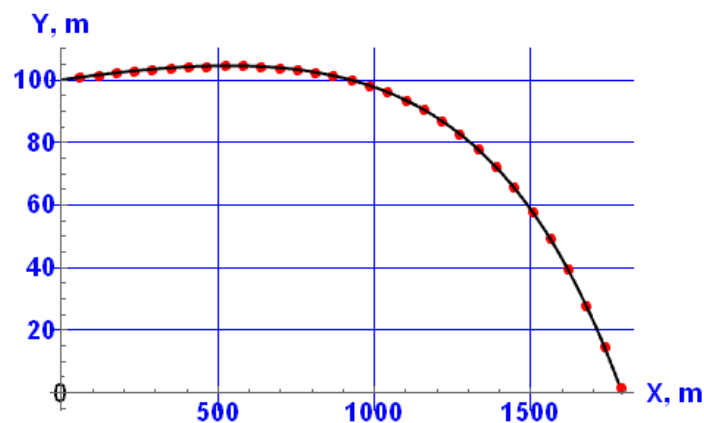

**Fig. 3**. Projectile trajectory at the height of the point of throw $y_0 = 100$ m.

The article [7] considers some aspects of the motion of table tennis balls as projectiles. The author notes the loss of accuracy of the results when using existing approximate analytical solutions. This is especially noticeable when table tennis balls move below the throwing plane and at large throwing angles. In order to verify these statements, with the help of formulas (4) dependences $y(x), x(t)$ are constructed for the values of the parameters [7] (see Fig. 4). As can be seen from Fig. 4, the analytical solution (4) describes the movement of table tennis balls with high accuracy. When applying formulas (4), no changes in the characteristics of the trajectories are observed. "Subtle features" [7] are mathematical phantoms resulting from the use of expansions of the true solution into truncated Taylor series some kind.

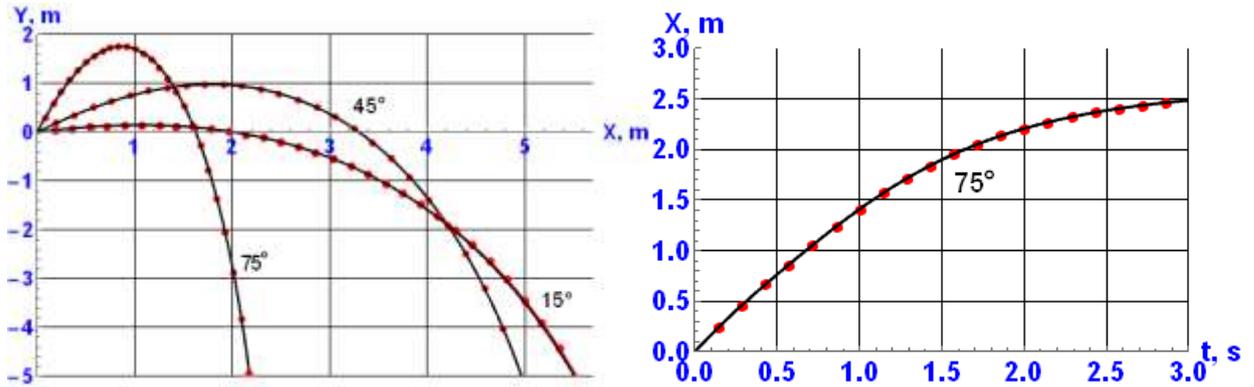

**Fig. 4.** Table tennis balls movement with parameters $V_0 = 7$ m/s, $g = 9.807$ m/s$^2$, $k = 0.01459$ s$^2$/m$^2$.

Of all the trajectories of sport projectiles, the trajectory of the shuttlecock has the greatest asymmetry. This is explained by the relatively large value of the drag coefficient $k$. In addition, the trajectory of the shuttle approaches to the vertical asymptote very quickly. In Fig.5 the value of the asymptote is determined by the numerical integration of the system of equations of motion of the projectile (2) and is $x_{as} = 8.67$ m. The asymptote value calculated using formula (5) is $x_{as} = 8.63$ m. The parameter calculation error is 0.5% in this case. Fig. 5 shows that the trajectory of the shuttlecock constructed using formulas (4) practically coincides with the asymptote. The author is not aware of any approximate analytical solution of the projectile motion problem that would describe the motion along the asymptote.

To check the applicability of formulas (4) at large throwing angles, we calculate the motion of a volleyball with the parameters [4]

$$V_0 = 14 \text{ m/s}, \quad \theta_0 = 85°, \quad k = 0.00369 \text{ s}^2/\text{m}^2, \quad g = 9.8 \text{ m/s}^2.$$

Note that in article [4], in order to construct solution (1), which agrees well with the numerical one, the value of $N = 120$ was used. This indicates a weak convergence of series (1). At the same time, the formulas (4) proposed here give excellent agreement with the numerical solution, which is illustrated in Fig. 6.

## 5. Conclusions

As examples of the use of formulas (4), the movement of various objects was considered - supersonic bullet, volleyball, shuttle of badminton, table tennis ball. The calculation results testify to the universality of the proposed analytical solutions (4), which are an improved version of formulas [8]. They are operable over a wide range of initial throw conditions and drag coefficients. The relative maximum deviation of the analytical value (4) from the numerical value (RK4) at any point of the trajectory does not exceed 1%. At very small values of the drag coefficient, formulas (4) are transformed into formulas of the theory of parabolic projectile

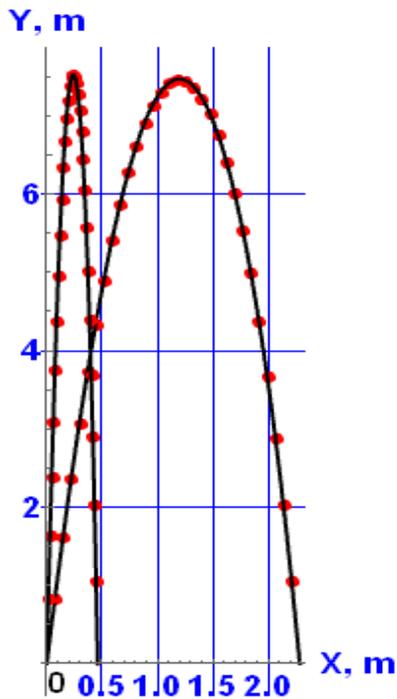 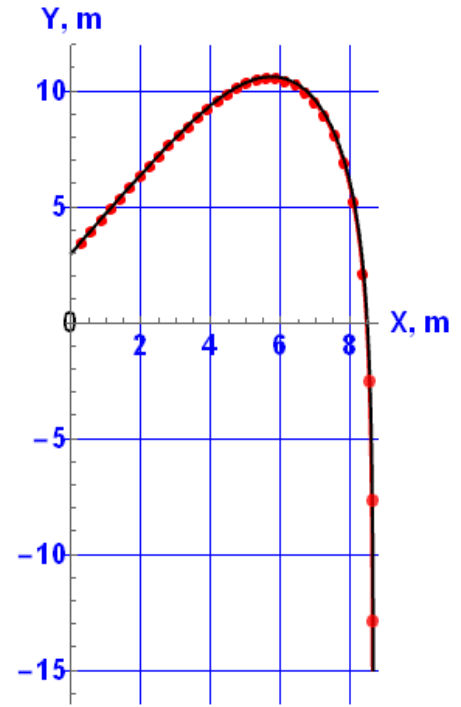

**Fig. 6.** Trajectory of volleyball at $\theta_0 = 85°, \theta_0 = 89°$.     **Fig.5.** Shuttlecock trajectory at

$$y_0 = 3\,m, \quad V_0 = 50\,m/s, \quad \theta_0 = 60°, \quad k = 0.022\,s^2/m^2, \quad g = 9.81\,m/s^2.$$

motion. It should be noted the educational and methodological benefits of solutions (4) for introductory physics courses studied by students of colleges and universities.

**References**

1. Turkyilmazoglu, M. (2016). Highly accurate analytic formulae for projectile motion subjected to quadratic drag. *European Journal of Physics*, *37*(3), 035001, https://doi.org/10.1088/0143-0807/37/3/035001

2. Belgacem, C. H. (2014). Range and flight time of quadratic resisted projectile motion using the Lambert W function. *European Journal of Physics*, **35**(5), 055025, https://doi.org/10.1088/0143-0807/35/5/055025

3. Belgacem, C.H. (2017). Analysis of projectile motion with quadratic air resistance from a nonzero height using the Lambert W function. *Journal of Taibah University for Science*, **11**(2), 328-391, http://doi.org/10.1016/j.jtusci.2016.02.009

4. Yabushita, K., Yamashita, M., Tsuboi, K. (2007). An analytic solution of projectile motion with the quadratic resistance law using the homotopy analysis method. *Journal of Physics A: Mathematical and Theoretical*, **40**, 8403-8416, http://doi:10.1088/1751-8113/40/29/015

5. Ray, S., Fröhlich, J. (2015). An analytic solution to the equations of the motion of a point mass with quadratic resistance and generalizations. *Archive of Applied Mechanics*, **85**(4), 395-414, https://doi.org/10.1007/s00419-014-0919-x

6. Das, C., Roy, D. (2014). Projectile motion with quadratic damping in a constant gravitational field. *Resonance*, **19**(5), 446-465, https://doi.org/10.1007/s12045-014-0048-4

7. Corvo, A. (2022). Subtle features in projectile motion with quadratic drag found through Taylor series expansions. *American Journal of Physics*, **90**(2), 135- 140,    https://doi.org/10.1119/10.0009227

8. Chudinov, P., Eltyshev, V.,  Barykin, Y. (2021). Analytical construction of the projectile motion trajectory in midair. *Momento Revista de Fisica*, **(62),** 79-96,  https://doi.org/10.15446/mo.n62.90752

9. Timoshenko, S. , Young, D (1948).   *Advanced Dynamics,*  McGraw-Hill Book Company, New-York, p. 112